\documentclass[12pt]{article}
\usepackage{amssymb, amsmath, amsthm, epsf, epsfig, setspace}
\newtheorem{theorem}{Theorem}[section]

\newtheorem{remark}[theorem]{Remark}
\newtheorem{lemma}[theorem]{Lemma}
\newtheorem{corollary}[theorem]{Corollary}

\title{Matrix Models for Beta Ensembles}  
\author{Ioana Dumitriu\thanks{Department of Mathematics, room 2-331, Massachusetts Institute of Technology, 77 Massachusetts Avenue, Cambridge, MA 02139 (dumitriu@math.mit.edu)} \hspace{.05cm} and Alan Edelman\thanks{ Department of Mathematics, room 2-388, Massachusetts Institute of Technology, 77 Massachusetts Avenue, Cambridge, MA 02139 (edelman@math.mit.edu; http://math.mit.edu/\~{}edelman)}}

\begin{document}
\maketitle
\abstract{This paper constructs tridiagonal random matrix models for
general ($\beta>0$) $\beta$-Hermite (Gaussian) and $\beta$-Laguerre
(Wishart) ensembles. These generalize the well-known Gaussian and Wishart models for $\beta = 1,2,4$. Furthermore, in the cases of the $\beta$-Laguerre ensembles, we eliminate the exponent quantization present in the previously known models.

We further discuss applications for the new matrix models, and present some open problems.}


\vspace{8.5cm}

\pagebreak

\section{Introduction} \label{intro}
\subsection{Overview}
Classical Random Matrix Theory focuses on the random matrix models in the following $3 \times 3$ table: 

\vspace{.3cm}

\begin{tabular}{l||cccccc}
&& \textbf{Real, $~\beta = 1$} && \textbf{Complex, $~\beta = 2$} && \textbf{Quaternion, $~\beta = 4$} \\ \hline \hline
\textbf{Hermite} && GOE && GUE && GSE \\
\textbf{Laguerre} && Real Wishart && Complex Wishart && (Quaternion Wishart) \\
\textbf{Jacobi}   && Real MANOVA  && Complex MANOVA && (Quaternion MANOVA) 
\end{tabular}

\vspace{.3cm}

The two entries in parentheses (in the third column) correspond to
less-studied random matrix models; the others are mainstream and have
been extensively researched and publicized. The three columns correspond
to Dyson's ``threefold way'' $\beta = 1,2$, and $4$; the three
rows correspond to the weight function associated to the random matrix model. Other weight functions have also been considered (for example, the uniform weight on the unit circle corresponds to the circular ensembles).

Zirnbauer \cite{Zirnbauer_10fold} and Ivanov \cite{ivanov} produced a more general taxonomy of random matrix models. Their characterizations (``tenfold'', 
respectively ``twelvefold'') are based on symmetric spaces, and 
include Hermite, Laguerre, and Jacobi cases, and also the circular
ensembles (each of their models can be associated with $\beta=1,2$ or $4$).

We propose a random matrix program of study that would generalize
$\beta$ beyond the above threefold way, thus generalizing the $3
\times 3$ cartesian product to $3 \times \infty$, making the
leap from discrete characterizations to continuous ones. A step in this direction has been initiated by Forrester \cite{Forrester_poly}, \cite{Forrester_book}, 
who studied the $\beta$-ensembles in connection with multivariate orthogonal polynomials and Calogero-Sutherland-type quantum systems. Furthermore, in the case of the classical Laguerre and Jacobi models, our program goes beyond the quantized exponents forced by the classical models, and proposes continuous ones.

\begin{center} 
\begin{figure}
\vspace{-5.5cm} 

\hspace{-2.5cm} \epsfig{figure = 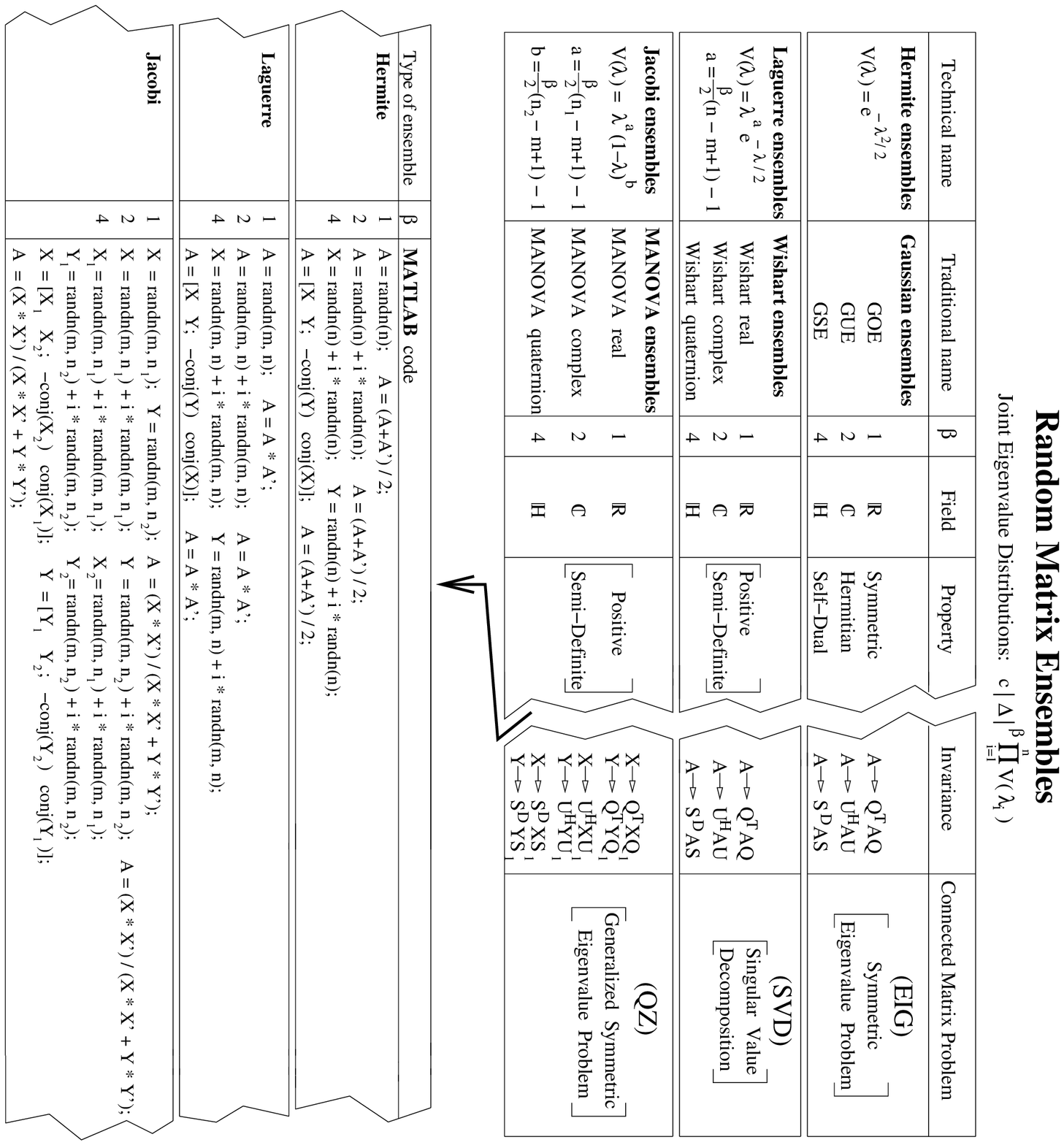} 

\vspace{-2cm}

\caption{Random Matrix Ensembles. As a guide to MATLAB notation, randn$(m,n)$ produces an $m \times n$ matrix with i.i.d.\ standard normal entries, conj$(X)$ produces the complex conjugate of the matrix $X$, and the $'$ operator produces the conjugate transpose of a matrix. Also $[X ~Y; ~Z ~W]$ produces a $2\times 2$ block matrix.}

\label{pretty_pic}
\end{figure}
\end{center}

For the benefit of the reader we have expanded the $ 3 \times 3$ table with detailed information in Figure~\ref{pretty_pic}.

\subsection{Background}

The Gaussian (or Hermite) ensembles arise in physics, and are identified by Dyson \cite{Dyson_3fold} by the group over which they are invariant:
Gaussian Orthogonal or for short GOE (with real 
entries), Gaussian Unitary or GUE (with complex entries), and Gaussian 
Symplectic or GSE (with quaternion entries). The Wishart ensembles arise 
in statistics, and the three corresponding models could be named Wishart 
real, Wishart complex, and Wishart quaternion.

The three Gaussian Ensembles have joint eigenvalue probability density function
\begin{eqnarray} \label{uno}
\mbox{HERMITE:~~} ~~~~&~& f_{\beta}(\lambda) = c_{H}^{\beta} \prod_{i<j} |\lambda_i - \lambda_j|^{\beta}
e^{-\sum_{i=1}^n \lambda_i^2/2}~, ~~~~~~~~~~~~~~~~~~~~~~~~~~~~~~
\end{eqnarray}
with $\beta =1$ corresponding to the reals, $\beta =2$ to the
complexes, $\beta = 4$ to the quaternions, and with 
\begin{eqnarray} \label{constH}
c_{H}^{\beta} = (2 \pi)^{-n/2} \prod_{j=1}^n \frac{\Gamma(1+\frac{\beta}{2})}{\Gamma(1+\frac{\beta}{2} j)}~.
\end{eqnarray}
The best references are Mehta 
\cite{mehta_book} and the original paper by Dyson \cite{Dyson_3fold}. 

Similarly, the Wishart (or Laguerre) models have joint eigenvalue p.d.f. 
\begin{eqnarray} \label{due}
\mbox{LAGUERRE:} &~~~~& f_{\beta}(\lambda) = c_{L}^{\beta,a} \prod_{i<j} |\lambda_i - \lambda_j|^{\beta} \prod_{i}
\lambda_i^{a-p} e^{-\sum_{i=1}^n\lambda_i/2}~,~~~~~~~~~
\end{eqnarray}
with $ a = \frac{\beta}{2}n$ and $p = 1+\frac{\beta}{2}(m-1)$. Again,
$\beta = 1$ for the reals, $\beta = 2$ for the complexes, and $\beta = 4$ for the quaternions. The constant 
\begin{eqnarray} \label{constL}
~~~~~~~~~~~~~~~~~~~~~c_{L}^{\beta, a} = 2^{-ma} \prod_{j=1}^{m}
\frac{\Gamma(1+\frac{\beta}{2})}{\Gamma(1+\frac{\beta}{2}j) \Gamma(a -
\frac{\beta}{2}(m-j)))}~.
\end{eqnarray}
Good references are \cite{muirhead82a}, \cite{edelman89a}, and \cite{james64a}, and for $\beta = 4$, \cite{MacDonald_book}.

To complete the triad of classical 
orthogonal polynomials, we will mention the $\beta$-MANOVA ensembles, 
which are associated to the multivariate analysis of variance (MANOVA) model. 
They are better known in the literature as the Jacobi ensembles, with joint eigenvalue p.d.f.
\begin{eqnarray} \label{tres}
~~~\mbox{JACOBI:} &~~~~~~~~~~& f_{\beta}(\lambda) =c_{J}^{\beta, a_1, a_2}
\prod_{i<j} |\lambda_i - \lambda_j|^{\beta} \prod_{j=1}^n
\lambda_i^{a_1 -p} ~
(1-\lambda_i)^{a_2 -p}~,
\end{eqnarray}
with $a_1 = \frac{\beta}{2}n_1$, $a_2 = \frac{\beta}{2}n_2$, and $p =
1+\frac{\beta}{2}(m-1)$. As usual, $\beta = 1$ for real and $\beta =
2$ for complex; also
\begin{eqnarray} \label{constJ}
~~~~~~~~~~~~~~~~~~~~~~~~~~~~~~~~~~c_{J}^{\beta, a_1, a_2} = \prod_{j=1}^m \frac{\Gamma(1+\frac{\beta}{2}) \Gamma(a_1+a_2-\frac{\beta}{2} (m-j))}{\Gamma(1+\frac{\beta}{2}j) \Gamma(a_1-\frac{\beta}{2}(m-j)) \Gamma(a_2-\frac{\beta}{2}(m-j))}~.
\end{eqnarray}
The MANOVA real and complex cases ($\beta=1$ and $2$) have been studied by statisticians (see \cite{muirhead82a}).

Though ``Gaussian'', ``Wishart'', and  ``MANOVA'' are the traditional names for 
the three types of $\beta$-ensembles, we prefer the sometimes used and technically more informative names ``Hermite'', ``Laguerre'', and ``Jacobi'' ensembles. 
These technical names reflect the fact that the p.d.f.'s for the 
ensembles correspond to the p.d.f.'s $\mbox{etr}(-A^2/2)$, $\mbox{det}(A)^{a-p}\mbox{etr}(-A/2)$, and $\mbox{det}(A)^{a_1-p} ~\mbox{det}(I-A)^{a_2-p}$ over their respective spaces of matrices. In turn, these functions correspond to three sets of orthogonal
polynomials (Hermite, Laguerre, Jacobi). Throughout this paper, we will use the term ``general $\beta$-Hermite, -Laguerre, -Jacobi ensembles'' for general $\beta$ in the p.d.f.s (\ref{uno}), (\ref{due}), (\ref{tres}).

Though it was believed that no other choice of $\beta$ would correspond to a matrix model constructed with entries from a classical distribution, there have been studies of general $\beta$-Hermite ensembles as theoretical eigenvalue distributions. They turn out to have important applications in lattice gas theory (see \cite{Forrester_book}, \cite{Forrester_poly}). 

The general $\beta$ ensembles appear to be connected to a broad 
spectrum of mathematics and physics, among which we list lattice gas 
theory, quantum mechanics, and Selberg-type integrals. Also, the $\beta$ 
ensembles are connected to the theory of Jack polynomials (with the 
correspondence $\alpha=\frac{2}{\beta}$ where $\alpha$ is the Jack 
parameter), which are currently objects of intensive research (see 
\cite{Stanley_jacks}, \cite{MacDonald_book}, \cite{Ok_Osh_shifted}).

\subsection{Our Results}

Dyson's original threefold way is a byproduct of the invariance assumptions as in the ``Invariance'' column of Figure \ref{pretty_pic}. By necessity, any invariant distribution is generically dense. Further, the invariance approach forces the consideration of the complex and quaternion division algebras.

In this paper, we drop the invariance requirement. What we gain are
``sparse'' models (with only O$(n)$ non-zero parameters) over the
reals numbers \emph{only}. As an additional bonus, we go beyond the
quantizations of the classical cases $\beta = 1,2,4$ and obtain
continuous exponents (see Section IV for further discussion of this point).

We provide real tridiagonal random matrix models for all
$\beta$-Gaussian (or Hermite) and $\beta$-Wishart (or Laguerre)
ensembles, and we discuss the possibility of constructing a real matrix model for the $\beta$-MANOVA (or Jacobi) ensembles. 

We obtain our results by extrapolating the classical cases,
thereby providing concrete models for what have 
previously been considered purely theoretical distributions. 

In Section II we establish results for symmetric 
tridiagonal matrices, and we use them to construct tridiagonal models for 
the $\beta$-Hermite ensembles. Along the way, we obtain a short proof
based on Random Matrix Theory for the Jacobian of the transformation
$T \rightarrow (q, \lambda)$, where $T$ is a symmetric tridiagonal
matrix, $\lambda$ is its set of eigenvalues, and $q$ is the first row
of its eigenvector matrix. In Section III we construct tridiagonal 
models for the $\beta$-Laguerre ensembles, by 
building on the same set of ideas that we use in Section II. In Section
IV we present some immediate applications of the new classes of ensembles 
and we discuss the $\beta$-Jacobi ensembles and other interesting open problems. 

We display our random matrix constructions in Table \ref{table}.

\begin{table}[ht]
\begin{center}
\begin{tabular}{|l||l|}
\hline 
& \\
$\begin{array}{c} \mbox{\textbf{Hermite} Matrix} \\ n \in \mathbb{N} \end{array}$ &  {\huge $H_{\beta}$} $ \sim  \frac{1}{\sqrt{2}} \left( \begin{array}{ccccc} N(0, 2) & \chi_{(n-1) \beta} & & & \\  \chi_{(n-1) \beta} & N(0, 2)  & \chi_{(n-2) \beta} & & \\  & \ddots & \ddots & \ddots & \\ & & \chi_{2\beta} & N(0,2) & \chi_{\beta} \\ & &  & \chi_{\beta} & N(0,2) \end{array} \right)$ \\
& \\
\hline
& \\
\textbf{Laguerre} Matrix 
& {\huge $L_{\beta}$}  $= B_{\beta}^{} B_{\beta}^{T}$, where \\
$\begin{array}{c} m \in \mathbb{N} \\ a \in \mathbb{R} \\
a>\frac{\beta}{2}(m-1) \end{array}$ & \hspace{1cm} $B_{\beta} \sim \left( \begin{array} {cccc} \chi_{2a} & & & \\ \chi_{\beta (m-1)} & \chi_{2a-\beta} & & \\ & \ddots & \ddots & \\ & & \chi_{\beta} & \chi_{2a-\beta(m-1)} \end{array} \right)$ \\
& \\
\hline
\end{tabular}
\end{center}
\caption{Random Matrix Constructions}
\label{table}
\end{table}

\section{The $\beta$-Hermite (Gaussian) Ensembles}

\subsection{Motivation: Tridiagonalizing the GOE, GUE, and GSE}

The joint distribution $f_\beta(\lambda)$ of the eigenvalues for the
GOE, GUE, and GSE  is
\begin{eqnarray} \label{once_again}
f_\beta(\Lambda) = c_{H}^{\beta} |\Delta(\lambda)|^{\beta} e^{-\frac{1}{2} \sum_i \lambda_i^2},
\end{eqnarray}
where $\beta=1,2,4$ \cite{mehta_book}. Here the Vandermonde determinant
notation $\Delta(\lambda)$ stands for~$\prod_{i \neq j}(\lambda_i-\lambda_j)$, and $c_{H}^{\beta}$ is given by (\ref{constH}).

We will prove in Section II.B that the tridiagonal $\beta$-Hermite random matrix displayed in Table \ref{table} has the joint eigenvalue p.d.f. given by general $\beta$ in (\ref{once_again}). For motivation, we will begin with a quick ``back-door'' proof for $\beta=1$ by tridiagonalizing the GOE; then we will extend the result to the GUE and GSE. 

To illustrate the proof and help the reader follow it more easily, we have included the diagram of Figure \ref{diagram}. 

\begin{theorem} \label{first_theorem} If A is an $n \times n$ matrix from the GOE, then 
reduction of $A$ to tridiagonal form shows that the matrix $T$ from
the $1$-Hermite ensemble has joint eigenvalue p.d.f. given by
(\ref{once_again}) with $\beta = 1$.
\end{theorem}

\begin{proof} We write $A = \left(\begin{array}{cc} a_n & x^{T} \\ x &
B \end{array} \right)$. Here $a_n$ is a standard Gaussian, $x$ is a
vector of $(n\!-\!1)$ i.i.d.\ Gaussians of mean 0 and variance $1/2$, and $B$
is an $(n\!-\!1)\times (n\!-\!1)$ matrix from the GOE; $a_n$, $x$ and $B$ are
all independent from each other.

Let $H$ be any $(n\!-\!1)\times(n\!-\!1)$ orthogonal matrix (depending only on $x$) such that 
\[
Hx ~=~ [ ||x||_2 ~0 ~\ldots ~0]^{T} ~\equiv~ ||x||_2 ~e_1~,
\]
where $e_1 = [1, 0, \ldots,0]^{T}$. Then clearly
\[
\left( \begin{array}{cc} 1 & 0 \\ 0 & H \end{array} \right ) ~\left( \begin{array}{cc} a_n & x^{T} \\ x & B \end{array} \right) ~ \left( \begin{array}{cc} 1 & 0 \\ 0 & H^{T} \end{array} \right ) ~~ = ~~ \left( \begin{array}{cc}  a_n & ||x||_2~ e_1^{T} \\ ||x||_2~ e_1 & HBH^{T} \end{array} \right)~.
\]

Since $A$ is from the GOE and $H$ depends only on $x$, we can readily identify the distributions of $a_n$, $||x||_2$ and $HBH^{T}$ (these three quantities are clearly independent). The entry $a_n$ is unchanged and thus a standard normal with variance 1. Being the length of a multivariate Gaussian of mean 0 and entry variance 1/2, $||x||_2$ has the distribution $\frac{1}{\sqrt{2}}\chi_{n-1}$. It is worth mentioning that the p.d.f. of $||x||_2$ is given by $\frac{2}{\Gamma(\frac{n-1}{2})}y^{n-2} e^{-y^2}$. Finally, by the orthogonal invariance of the GOE, $HBH^{T}$ is an $(n\!-\!1) \times (n\!-\!1)$ matrix from the GOE.

Proceeding by induction completes the tridiagonal construction.

Because the only operations we perform on $A$ are orthogonal similarity
transformations, which do not affect the eigenvalues, the conclusion
of the theorem follows.
\end{proof}

We recall that matrices from the GOE have the following properties:

\textbf{Property 1.} the joint eigenvalue density is 
$c_{H}^{1}~|\Delta(\lambda)| e^{-\frac{1}{2} \sum\limits_i
\lambda_i^2}$ \cite{mehta_book};

\textbf{Property 2.} the first row of the eigenvector matrix is distributed uniformly on the sphere, and it is independent of the eigenvalues.

\vspace{.25cm}

The second property is an immediate consequence of the fact that the
eigenvector matrix of a GOE matrix is independent from the
eigenvalues \cite[(3.1.3) and (3.1.16), pages 55-58]{mehta_book}, and has the Haar
(uniform) distribution because of the orthogonal invariance. 

The following corollary is easily established.

\begin{corollary} If $T$ is a matrix from the $1$-Hermite ensemble, with eigendecomposition $T = Q\Lambda Q^{T}$, then the first row $q$ of the eigenvector matrix $Q$ is independent of $\Lambda$, and is distributed uniformly on the sphere.
\end{corollary}

\begin{proof}
If $A = Q_1 \Lambda Q_1^{T}$ and $T = HAH^{T}$, then $Q = HQ_1$. Since each one of the reflectors which form H has first row $e_1$, multiplication by $H$ does not affect the first row of $Q_1$. The conclusion follows. 
\end{proof}

Reduction to tridiagonal form is a familiar algorithm which solves the symmetric eigenvalue problem. The special ``reflector'' matrix $H$ used in practice for a vector $x = [x_1, \ldots, x_{n-1}]^T$ is 
\[
H = I - 2 \frac{uu^{T}}{u^{T}u}~,
\]
where $u = x \pm x_1~e_1$. This special matrix $H$ is known as the ``Householder reflector'' (see \cite[page 209]{golub-vanloan}).

 The tridiagonal reduction algorithm can be applied to any real symmetric, complex hermitian, or quaternion self-dual matrix; the resulting matrix is always a real, symmetric tridiagonal. Using the algorithm similarly on a GUE or GSE matrix one gets the following 

\begin{corollary} When $\beta = 2,4$, reduction to tridiagonal form of
matrices from the $GUE$, respectively $GSE$, shows that the
tridiagonal $2$-Hermite, respectively $4$-Hermite, random matrix has the distribution given by (\ref{once_again}). Note that $\beta$ ``counts'' the number of independent Gaussians in each entry of the matrix.
\end{corollary}

\begin{remark} 
The observation that Numerical Linear Algebra algorithms may be performed statistically is not new; it may be found in the literature (see Trotter \cite{trotter84a}, Silverstein \cite{silverstein85a}, Edelman \cite{edelman89a}). 
\end{remark}

\subsection{Tridiagonal Matrix Lemmas}

In this section we prove lemmas that will be used in our constructions in Sections II.C and III.B. 

Given a tridiagonal matrix $T$ defined by the diagonal $a = (a_n,
\ldots, a_1)$ and sub-diagonal $b = (b_{n-1}, \ldots, b_1)$, with
all $b_i$ positive, let $T = Q \Lambda Q^{T}$ be the
eigendecomposition of $T$ as in Theorem~\ref{main1}. 
Let $q$ be the first row of $Q$ and $\lambda = $ diag $(\Lambda)$.

\begin{lemma} Under the assumptions above, starting from $q$ and $\lambda$, one can uniquely reconstruct $Q$ and $T$. 
\end{lemma}
\begin{proof}
This is a special case of the more general Theorem 7.2.1 in Parlett \cite{Parlett_book}.
\end{proof}

\begin{remark} \label{bijection}
It follows that, except for sets of measure 0, the map $T \rightarrow
(q, \lambda)$ is a bijection from the set of tridiagonal matrices of
size $n$ with positive sub-diagonal, to the set of pairs $(q,
\lambda)$, with $q$ a unit norm $n$-dimensional vector of positive real entries, and $\lambda$ a strictly increasingly ordered sequence of $n$ real numbers. Let the bijection's Jacobian be denoted by $J$ $\left(J = \left \{ \frac{\partial{(a,b)}}{\partial{(q,\lambda)}} \right \} \right)$.
\end{remark}

Our next lemma establishes a formula for the Vandermonde determinant of the eigenvalues of a tridiagonal matrix.

\begin{lemma} \label{one} The Vandermonde determinant for the ordered
eigenvalues of a symmetric tridiagonal matrix with positive
sub-diagonal $b = (b_{n-1}, \ldots, b_1)$ is given by 
\[
\Delta(\lambda) =  \prod_{i<j} (\lambda_i - \lambda_j) = \frac{\prod\limits_{i=1}^{n-1} b_i^i}{\prod\limits_{i=1}^n q_i}~,
\]
where $(q_1, \ldots, q_n)$ is the first row of the eigenvector matrix.  
\end{lemma}

\begin{proof} Let $\lambda_i^{(k)},~i=1 \ldots k$, be the eigenvalues of the $k \times k$ lower right-corner submatrix of $T$. Then $P_{k}(x) = \prod_{i=1}^k (x-\lambda_i^{(k)})$ is the associated characteristic polynomial of that submatrix. 

For $k=1, \ldots, n$ we have the three-term recurrence
\begin{eqnarray} \label{unu}
P_{k}(x) = (x-a_k) P_{k-1}(x) - b_{k-1}^2 P_{k-2}(x)~,
\end{eqnarray}
and the two-term relation
\begin{eqnarray} \label{doi}
\prod_{\begin{array}{c}1 \leq i \leq k \\ 1 \leq j \lq k-1 \end{array}} |\lambda_i^{(k)} - \lambda_j^{(k-1)}| = \prod_{i=1}^k |P_{k-1} (\lambda_i^{(k)})| = \prod_{j=1}^{k-1} |P_{k}(\lambda_j^{(k-1)})|~.
\end{eqnarray}

{F}rom~(\ref{unu}) we get
\begin{eqnarray} \label{trei}
|\prod_{i=1}^{k-1} P_{k} (\lambda_i^{(k-1)})| = b_{k-1}^{2(k-1)} |\prod_{i=1}^{k-1} P_{k-2}(\lambda_i^{(k-1)})|~.
\end{eqnarray}
By repeatedly applying (\ref{unu}) and (\ref{two}) we obtain
\begin{eqnarray} \label{patru}
\prod_{i=1}^{n-1} |P_{n} (\lambda_i^{(n-1)})| & = & b_{n-1}^{2(n-1)} ~~~~~~~~~~~~~~\prod_{i=1}^{n-2} |P_{n-1}(\lambda_i^{(n-2)})| \\
& = & b_{n-1}^{2(n-1)} ~ ~b_{n-2}^{2(n-2)} ~~|\prod_{i=1}^{n-2} P_{n-3}(\lambda_i^{(n-2)})| \\
& = &  \ldots \\
& = & \prod\limits_{i=1}^{n-1} b_i^{2i}~.
\end{eqnarray}

Finally, we use the following formula due to C.C. Paige, found in \cite{Parlett_book} as the more general Theorem 7.9.2:
\begin{eqnarray} \label{cinci}
q_i^2 = \left | \frac{P_{n-1}(\lambda_i)}{P_{n}'(\lambda_i)} \right |=  \left |\frac{P_{n-1}(\lambda_i^{(n)})}{P_{n}'(\lambda_i^{(n)})} \right |~.
\end{eqnarray}
It follows that
\begin{eqnarray} \label{sase}
\prod_{i=1}^n q_i^2 = \frac{\prod\limits_{i=1}^n |P_{n-1}(\lambda_i^{(n)})|}{\Delta(\lambda)^2} = \frac{\prod\limits_{i=1}^{n-1} b_i^{2i}}{\Delta(\lambda)^2}~,
\end{eqnarray}
which proves the result.
\end{proof}

\begin{remark} The Vandermonde determinant formula of Lemma \ref{one}
can also be obtained from the Heine formula, as presented in Deift \cite{deift_book}, page 44.
\end{remark}

\begin{figure}[ht]
\begin{center}
\epsfig{figure = 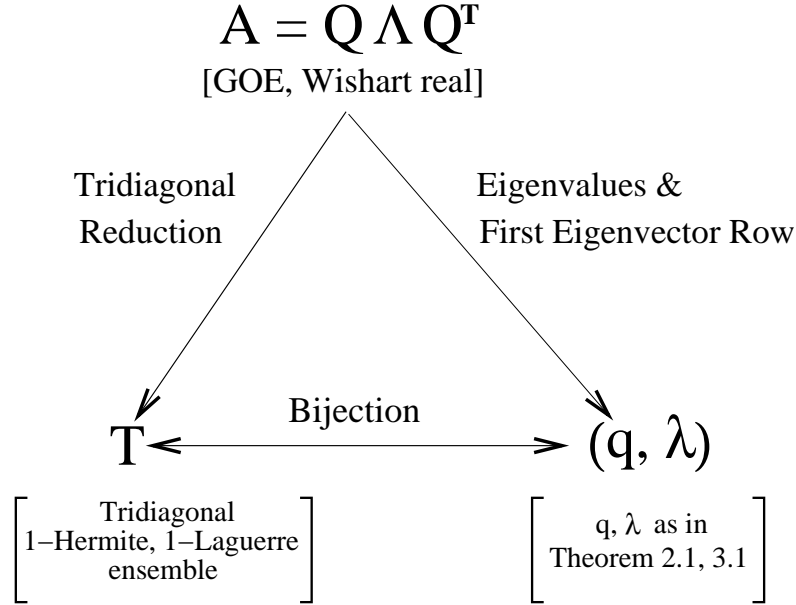, height = 8cm}
\caption{A dense symmetric matrix $A$ can be tridiagonalized (left side) or diagonalized (right side). In brackets, we provide the distributions starting with that of $A$ (GOE or Wishart real).}

\label{diagram}
\end{center}
\end{figure}

The next lemma section computes the Jacobian $J$ by relating the tridiagonal and diagonal forms of a GOE matrix, as in Figure \ref{diagram}.

\begin{lemma} \label{two} The Jacobian $J$ can be written as
\[
J~ =~\frac{\prod\limits_{i=1}^{n-1} b_i}{\prod\limits_{i=1}^n q_i}~.
\]
\end{lemma}

\begin{proof}
To obtain the Jacobian, we will study the transformation from GOE to $1$-Hermite ensemble (see Figure \ref{diagram}). Note that $J$ does \emph{not} depend on $\beta$; hence computing the Jacobian for this case is sufficient.

Let $T$ be a $1$-Hermite matrix. We know from Section II.A that the eigenvalues of $T$ are distributed as the eigenvalues of a symmetric GOE matrix $A$, from which $T$ can be obtained via tridiagonal reduction ($T = HAH^{T}$ for some orthogonal $H$, which is the product of the consecutive reflections described in Section II.A). 

The joint element distribution for the matrix $T$ is 
\[
\mu(a, b)~=~ c_{a,b}~e^{-\frac{1}{2} \sum\limits_{i=1}^n a_i^2} \prod_{i=1}^n b_i^{i-1} e^{-\sum\limits_{i=1}^n b_i^2}~~,~~~~~~~\mbox{where}~~~~~~~~~~~c_{a,b} = \frac{2^{n-1}}{(2 \pi)^{n/2} \prod\limits_{i=1}^{n-1} \Gamma(\frac{i}{2})}~. 
\]

Let 
\[
da = \wedge_{i=1}^n da_i,~~ db = \wedge_{i=1}^{n-1} db_i, ~~d\lambda =
 \wedge_{i=1}^n \lambda_i,~
\]
and $dq$ be the surface element of the $n$-dimensional sphere. 
Let $\mu(a(q, \lambda), b(q, \lambda))$ be the expression for 
$\mu(a,b)$ in the new variables $q, \lambda$. We have that 
\begin{eqnarray} \label{measure_change}
\mu \left(a,b \right)~da~db ~= ~J~\mu \left(a(q, \lambda), b(q, \lambda)\right)~dq~d\lambda~ ~\equiv ~ \nu(q, \lambda) ~dq~d\lambda.
\end{eqnarray}

We combine Properties 1 and 2 of Section II.A to get the joint p.d.f. $\nu(q,\lambda)$ of the eigenvalues and first eigenvector row of a GOE matrix, and rewrite it as
\[
\nu(q, \lambda)~dq~d\lambda ~=~ n!~c_{H}^1 ~\frac{2^{n-1} \Gamma(\frac{n}{2})}{\pi^{n/2}}~\Delta(\lambda) ~e^{-\frac{1}{2} \sum\limits_i \lambda_i^2}~dq~d\lambda.
\]
We have introduced the $n!$ and removed the absolute value from the
Vandermonde, because the eigenvalues are ordered. We have also
included the distribution of $q$ (as mentioned in Property 2, it is
uniform, but only on the all-positive $2^{-n}$th of the sphere
because of the condition $q_i \geq 0$).

Since orthogonal transformations do not change the Frobenius norm $||A||_F =
\sum\limits_{i,j=1}^n a_{ij}^2$ of a matrix $A$, from (\ref{measure_change}), it follows that
\[
J ~= ~\frac{\nu(q, \lambda)}{\mu(a,b)} = \frac{n! ~c_{H}^1 \frac{2^{n-1} \Gamma(\frac{n}{2})}{\pi^{n/2}}}{c_{a,b}}~\frac{\Delta(\lambda)}{\prod\limits_{i=1}^n b_i^{i-1}}~.
\]

All constants cancel, and by Lemma~\ref{one} we obtain
\[
J ~= ~\frac{\prod\limits_{i=1}^{n-1} b_i}{\prod\limits_{i=1}^n q_i}~.
\]

Note that we have not expressed $\mu(a,b)$ in terms of $q$ and $\lambda$ in the above, and have thus obtained the expression for the Jacobian neither in the variables $q$ and $\lambda$, nor $a$ and $b$, solely; but rather in a mixture of the two sets of variables. The reason for this is that of simplicity.
\end{proof}

\begin{remark} Our derivation of the Jacobian is a true Random Matrix derivation. Alternate derivations of the Jacobian can be obtained
either via symplectic maps or through direct calculation. We thank Percy Deift and Peter Forrester, respectively, for having shown them to us.
\end{remark}

The last lemma of this section computes one more Jacobian, which will be needed in Section III.B. 

Let $B$ be a bidiagonal matrix with positive diagonal $x = (x_m, \ldots, x_{1})$ and positive sub-diagonal $y = (y_{m-1}, \ldots, y_1)$. Let $T = BB^T$; denote by $a = (a_{m}, \ldots a_1)$ and $b = (b_{m-1}, \ldots, b_1)$ the diagonal, respectively the sub-diagonal part of $T$. Since $T$ is a positive definite matrix, the transformation $B \rightarrow T$ is a bijection from the set of bidiagonal matrices with positive entries to the set of positive definite tridiagonal matrices.

\begin{lemma} \label{three} The Jacobian $J_{(B \rightarrow T)}$ is
\[
J_{(B \rightarrow T)} = \Big(2^{m} x_{1} \prod_{i=2}^{m} x_{i}^2 \Big)^{-1}~.
\]
\end{lemma}

\begin{proof} We compute $J_{(B \rightarrow T)}$ from the formula
\[
dx~dy = J_{(B \rightarrow T)} ~da~db~,
\]
where $dz = \wedge_{i} dz_i$ for all $z \in \{a,b,x,y\}$.

We have that 
\begin{eqnarray}
a_m & = & x_m^2~, \\
a_i & = & y_{i}^2+x_{i}^2~, \\
b_i & = & y_i x_{i+1}~, 
\end{eqnarray}
for all $i = m-1, m-2, \ldots, 1$. 

Hence by computing differentials we get
\begin{eqnarray*} 
da_m & = & 2x_m ~dx_m \\
\label{un} da_i & = & 2(x_{i}~dx_{i} + y_i ~dy_i), ~\forall i=m-1,m-2,\ldots,1 \\
\label{deux} db_i & = & x_{i+1}~dy_i+y_i~dx_{i+1}, ~\forall i=m-1,m-2,\ldots,1~,
\end{eqnarray*}
from which the formula follows.

\end{proof}

\subsection{The Eigendistribution of the $\beta$-Hermite ensemble}

Let $H_{\beta}$ be a random real symmetric, tridiagonal matrix whose distribution we schematically depict as
\[ H_{\beta} \sim  \frac{1}{\sqrt{2}} \left( \begin{array}{ccccc} N(0, 2) & \chi_{(n-1) \beta} & & & \\  \chi_{(n-1) \beta} & N(0, 2)  & \chi_{(n-2) \beta} & & \\  & \ddots & \ddots & \ddots & \\ & & \chi_{2\beta} & N(0,2) & \chi_{\beta} \\ & &  & \chi_{\beta} & N(0,2) \end{array} \right)
\] 

By this we mean that the $n$ diagonal elements and the $n-1$ sub-diagonals are mutually independent, with standard normals on the diagonal, and $\frac{1}{\sqrt{2}} \chi_{k \beta}$ on the sub-diagonal. 

\begin{theorem} \label{main1} Let $H_{\beta} = Q \Lambda
Q^{T}$ be the eigendecomposition of $H_{\beta}$; fix the signs
of the first row of $Q$ to be non-negative and order the eigenvalues in increasing order on the diagonal of $\lambda = $ {\rm diag}$(\Lambda)$. Then $\lambda$ and $q$, the first row of $Q$, are independent. Furthermore, the joint density of the eigenvalues is 
\[
f_{\beta}(\lambda) = c_{H}^{\beta} ~\prod_{i<j} |\lambda_i - \lambda_j|^{\beta} ~e^{-\frac{1}{2} \sum\limits_{i=1}^n \lambda_i^2} ~~=~~ c_{H}^{\beta}~|\Delta(\lambda)|^{\beta} ~e^{-\frac{1}{2} \sum\limits_{i=1}^n \lambda_i^2} ~,
\]
and $q = (q_1, \ldots, q_n)$ is distributed as $(\chi_{\beta}, \ldots, \chi_{\beta})$, normalized to unit length. 
\end{theorem}

\textit{Proof of Theorem~\ref{main1}}. 

Just as before, we denote by $a = (a_{n}, \ldots, a_{1})$ the diagonal of $H_{\beta}$, and by $b = (b_{n-1}, \ldots, b_1)$ the sub-diagonal. The differentials $da, db, dq, d\lambda$ are the same as in Lemma \ref{two}.

For general $\beta$, we have that 
\[
(dH_{\beta}) \equiv \mu(a,b)~da~db = c_{a,b} ~ \prod_{k=1}^{n-1} b_k^{k\beta-1} e^{-\frac{1}{2} ||T_1||_F}~da~db = c_{a,b} ~ J~  \prod_{k=1}^{n-1} b_k^{k\beta-1} e^{-\frac{1}{2} ||T_1||_F} ~dq~d\lambda ~,
\]
where 
\[
c_{a,b} = \frac{2^{n-1}}{(2\pi)^{n/2} \prod_{k=1}^{n-1} \Gamma \left( \frac{\beta}{2}k \right)}~.
\]

With the help of Lemmas \ref{one} and \ref{two} this identity becomes 

\begin{eqnarray} 
(dH_{\beta})  & = & c_{a,b}~\frac{\prod_{k=1}^{n-1} b_k}{\prod_{k=1}^n q_k}  \prod_{k=1}^{n-1} b_k^{k\beta-1} e^{-\frac{1}{2} ||T_1||_F} ~dq ~d\lambda \\
\label{final_pdf} & = & c_{a,b}~\frac{\prod_{k=1}^{n-1} b_k^{k\beta}}{\prod_{i=1}^n q_i^{\beta}}  \prod_{i=1}^n q_i^{\beta-1} e^{-\frac{1}{2} \sum_i \lambda_i^2} ~dq ~d\lambda~.
\end{eqnarray}

Thus
\[
(dH_{\beta})  = \left(c_{q}^{\beta} \prod_{i=1}^n q_i^{\beta-1} ~dq \right)~
\left(n!~c_{H}^{\beta} ~\Delta(\lambda)^{\beta}  e^{-\frac{1}{2} \sum_i \lambda_i^2}
~d\lambda \right)~.
\]

Since the joint density function of $q$ and $\lambda$ separates, $q$ and $\lambda$ are independent. Moreover, once we drop the ordering imposed on the eigenvalues, it follows that the joint
eigenvalue density of $H_{\beta}$ is $c_{H}^{\beta}~ |\Delta(\lambda)|^{\beta}  e^{-\frac{1}{2} \sum_i \lambda_i^2}$, and $q$ is distributed as $(\chi_{\beta}, \ldots, \chi_{\beta})$, normalized to unit length. {F}rom (\ref{final_pdf}), it also follows that 
\begin{eqnarray} \label{constq}
c_{q}^{\beta} = \frac{2^{n-1} \Gamma(\frac{\beta}{2}n)}{\left[\Gamma(\frac{\beta}{2})\right]^n}~.
\end{eqnarray} 

\qed

\section{The $\beta$-Laguerre (Wishart) Ensembles}

\subsection{Motivation: Tridiagonalizing the Wishart ensembles}

The preceding section gives tridiagonal random matrix models for all
$\beta$-Hermite ensembles. In the following we define the
$\beta$-Laguerre ensembles, and give tridiagonal random matrix models for them.

The Wishart ensembles have joint eigenvalue density 
\begin{eqnarray} \label{quattro}
f_{\beta}(\lambda) = c_{L}^{\beta,a}~ |\Delta(\lambda)|^{\beta} ~\prod\limits_{i=1}^m \lambda_i^{a-p} e^{-\sum_{i=1}^m \lambda_i/2}~,
\end{eqnarray}
again with $a = \frac{\beta}{2}n$, $p = 1+\frac{\beta}{2}(m-1)$, and with $\beta = 1$ for real,
respectively $\beta = 2$ for complex. Here $c_{L}^{\beta,a}$ as the same as in (\ref{constL}).

From now on $p$ will always denote the quantity $1+\frac{\beta}{2}(m-1)$,
following the notation of Muirhead for $\beta =1$ \cite{muirhead82a} (chapter 7) and Forrester
\cite{Forrester_book}(Forrester uses $1+\frac{1}{\alpha}(m-1)$,
where $\alpha = 2/\beta$ is the Jack parameter). Its presence is
implicit in the p.d.f. of all $\beta$-Laguerre ensembles; hence we
will identify the ensembles by $\beta$ and by $a$ (we call the latter
the ``Laguerre'' parameter, generalizing from the univariate case
$\beta = 1$, $m=1$).

As in Section II.A, we will provide the most basic case for our
construction: the case $\beta = 1$ and Wishart real exponent
$\frac{n-m-1}{2}$ (also referred as the case $\beta = 1$ and Laguerre
parameter $a = \frac{n}{2}$).

\begin{theorem} Let $G$ be an $m \times n$ matrix of i.i.d.\ standard
Gaussians; then $W = GG^{T}$ is a Wishart real matrix. By reducing $G$
to bidiagonal form $B$ one obtains that the matrix $T= BB^{T}$ from the
$1$-Laguerre ensemble of Laguerre parameter $a = \frac{n}{2}$ (defined
as in Table \ref{table}) has the joint eigenvalue p.d.f. given by (\ref{quattro}).
\end{theorem}

\begin{proof} We write 
\[
G ~=~ \left( \begin{array}{c} x^{T} \\ G_1 \end{array}
\right)~,
\]
with $x^{T}$ a row multivariate standard Gaussian of length $n$ and $G_1$ a
$(m-1) \times n$ matrix of i.i.d.\ standard Gaussians. Let $R$ be
a right reflector corresponding to the vector $x^{T}$ ($R^{T} x =
||x||_2~e_1^{T}$) which is independent of $G_1$. Hence $G_1R$ is a matrix
of i.i.d.\ standard Gaussians. 

Write $G_1 R = [y, ~G_2]$, where $y$ is a column multivariate standard
Gaussian of length $m-1$ and $G_2$ is a $(m-1) \times (n-1)$ matrix of
i.i.d.\ standard Gaussians. Let $L$ be a left reflector corresponding
to $y$ ($Ly = ||y||_2 ~e_1$) which is independent of $G_2$. Then we
have that
\[
\left( \begin{array}{cc} 1 & 0 \\ 0 & L \end{array} \right) ~G~R ~~=~~
\left( \begin{array}{cc} ||x||_2 & 0 \\ ||y||_2 ~e_1 & LG_2
\end{array} \right)~.
\]
As we have seen before, $||x||_2$ is distributed like $\chi_{n-1}$,
$||y||_2$ is distributed like $\chi_{m-1}$, and $LG_2$ is a matrix of
i.i.d.\ standard Gaussians (since $L$ and $G_2$ are independent).

We proceed inductively to finish the bidiagonal construction of $B$.

Because the operations we have performed  on $G$ are orthogonal left
and right multiplications, which do not affect the singular values, it
follows that the singular values of $G$ and $B$ are the same. 
Since the squares of the singular values of $G$, 
respectively $B$, are the eigenvalues of $W$, respectively $T$, the conclusion of the theorem follows. 
\end{proof}

\begin{remark} The bidiagonalization process presented above is part of a familiar Numerical Linear Algebra algorithm for computing the singular values of a matrix.
\end{remark}

\begin{corollary}
The same process of bidiagonalization performed on $\tilde{G}, $, a
matrix of i.i.d.\ standard complex (standard quaternion) Gaussians, shows that the matrix
$\tilde{W} = \tilde{G}\tilde{G}^{T}$ and the matrix $T$ from the
$2$-Laguerre ($4$-Laguerre) ensemble of parameter $a = n$ ($a= 2n$) 
has the joint eigenvalue
p.d.f. given by (\ref{quattro}). In all three cases (real, complex,
quaternion) we say that $T$ represents the tridiagonalization of the
Wishart (real, complex, quaternion) ensemble.
\end{corollary}

In the next section we prove the general form of the theorem.

\subsection{The Eigendistribution of $\beta$-Laguerre Ensemble}

Let 
\[
B_{\beta}  \sim \left( \begin{array}{ccccc} \chi_{2a}  & & & & \\
				   \chi_{\beta(m-1)} & \chi_{2a-\beta}& & & \\
						& \ddots & \ddots &  &\\
						& & \chi_{\beta} &
				   \chi_{2a-\beta(m-1)}  & \end{array}
				   \right)~,
\]
by this meaning that all of the $2m-1$ diagonal and subdiagonal
elements are mutually independent with the corresponding $\chi$
distribution. 

Let $L_{\beta} = B_{\beta}^{} B_{\beta}^{T}$ be the corresponding tridiagonal matrix.

\begin{theorem}  \label{main2}
Let $L_{\beta} = Q \Lambda Q^{T}$ be the eigendecomposition of
$L_{\beta}$; fix the signs of the first row of $Q$ to be non-negative
and order the eigenvalues increasingly on the diagonal of
$\Lambda$. Then $\Lambda$ and the first row $q$ of $Q$ are independent. Furthermore, the joint density of the eigenvalues is 
\[
f_{\beta}(\lambda) = c_{L}^{\beta,a}~ |\Delta(\lambda)|^{\beta} ~\prod_{i=1}^n \lambda_i^{a-p} e^{-\sum_{i=1}^n \lambda_i/2}~,
\]
where $p = 1+\frac{\beta}{2}(m-1)$, and $q$ is distributed as $(\chi_{\beta}, \ldots, \chi_{\beta})$ normalized to unit length. 
\end{theorem}

\textit{Proof of Theorem \ref{main2}}. 
We will use throughout the results of Lemma \ref{one}, Lemma \ref{two}, Lemma \ref{three} and Remark \ref{bijection}, which are true in the context of tridiagonal symmetric matrices with positive sub-diagonal entries. By definition, $L_{\beta}$ is such a matrix.

We will again use the notations of Lemma \ref{two} and \ref{three} for the differentials $da$, $db$, $dq$, $d\lambda$, $dx$, and $dy$. 

We define $(dB_{\beta}$ to be the joint element distribution on
$B_{\beta}$
\begin{eqnarray*}
(dB_{\beta} ) ~\equiv ~\mu(x,y) ~dx~dy~ = ~\prod_{i=0}^{m-1}
x_{m-i}^{a-\beta i-1} e^{-x_{i}^2/2}~\prod_{i=1}^{m-1} y_i^{\beta i -1} e^{-y_i^2/2}~dx~dy ~.
\end{eqnarray*}

By using Lemma \ref{quattro} we obtain the joint element distribution on $L_{\beta}$ as
\begin{eqnarray} \label{jac} 
(dL_{\beta}) & \equiv & J_{B\rightarrow T}^{-1} \mu(x,y) ~dx~dy \\
\label{jac1} & = & 2^{-m} c_{x,y}~x_{1}^{2a-\beta(m-1)-2}  e^{-x_{1}^2/2} \prod_{i=0}^{m-2}
x_{m-i}^{a-\beta i-3} e^{-x_{i}^2/2}
\prod_{i=1}^{m-1} y_i^{\beta i -1} e^{-y_i^2/2}~dx~dy~,
\end{eqnarray}
where \[
c_{x,y} = \frac{\prod_{i=1}^{m-1} \Gamma(i\frac{\beta}{2}) \prod_{i=1}^m \Gamma(a-\frac{\beta}{2}(i-1))}{2^{2m-1}}~.
\]

We rewrite (\ref{jac1}) in terms of $x, y, \lambda,$ and $q$:
\begin{eqnarray*}
(dL_{\beta}) & = & 2^{-m} c_{x,y}   ~e^{-\sum\limits_{i=1}^{m} x_{i}^2/2} e^{-\sum\limits_{i=1}^{m-1} y_i^2/2} \frac{\prod_{i=1}^{m-1} (x_{i+1}
y_{i})}{\prod_{i=1}^{m} q_{i}}  x_{1}^{2a-\beta(m-1)-2}  \times \\
& & ~~~~~~~ \times \prod_{i=0}^{m-2} x_{m-i}^{2a-\beta (m-i)-3}
\prod_{i=1}^{m-1} y_i^{\beta i -1} ~dq~d\lambda\\
& = & 2^{-m} c_{x,y}   ~ e^{-\sum\limits_{i=1}^{m} x_{i}^2/2} e^{-\sum\limits_{i=1}^{m-1} y_i^2/2} \frac{\prod_{i=0}^{m-1}
x_{m-i}^{2a-\beta(m-i)-2} \prod_{i=1}^{m-1} y_{i}^{\beta i}}{\prod_{i=1}^{m}
q_i}~dq~d\lambda.
\end{eqnarray*}

Since the Vandermonde with respect to $b$ and $q$ and the ordered eigenvalues $\lambda$ can be written as
\[
\Delta(\lambda) = \frac{\prod_{i=1}^{m-1} b_i^i}{\prod_{i=1}^m q_i}~,
\]
it follows that 
\[
\Delta(\lambda) = \frac{\prod_{i=1}^{m-1} \Big(x_{i+1} y_{i}\Big)^{i}}{\prod_{i=1}^m q_i}~.
\]
This means that we can rewrite 
\begin{eqnarray*}
(dL_{\beta}) & = & 2^{-m} c_{x,y}   ~e^{-\sum\limits_{i=0}^{m-1} x_{n-i}^2/2} e^{-\sum\limits_{i=1}^{m-1} y_i^2/2} \frac{\prod_{i=1}^{m-1} \Big(x_{i+1} y_{i}\Big)^{\beta i}}{\prod_{i=1}^m q_i^{\beta}} \prod_{i=1}^{m-1} q_i^{\beta-1} \prod_{i=0}^{m-1} x_{m-i}^{2a-\beta(m-1)-2} ~dq~d\lambda\\
	& = & 2^{-m} c_{x,y}   ~e^{-\sum\limits_{i=0}^{m-1} x_{n-i}^2/2} e^{-\sum\limits_{i=1}^{m-1} y_i^2/2} \Delta(\lambda)^{\beta} \prod_{i=1}^{m-1} q_i^{\beta-1} \left(\prod_{i=0}^{m-1} x_{m-i} \right)^{2a-\beta(m-1)-2} ~dq~d\lambda~.
\end{eqnarray*}

The trace and the determinant are invariant under
orthogonal similarity transformations, so tr$(L_{\beta})=$ tr$(\Lambda)$, and det$(L_{\beta})=$ det$(\Lambda)$. This is equivalent to 
\begin{eqnarray*}
\sum_{i=0}^{m-1} x_{m-i}^2 + \sum_{i=1}^{m-1} y_i^2 & = & \sum_{i=1}^m \lambda_i~,\\
\prod_{i=0}^{m-1} x_{m-i}^2 & = & \prod_{i=1}^{m} \lambda_i~.
\end{eqnarray*}
Using this, and substituting $p$ for $1+\frac{\beta}{2}(m-1)$, we obtain that 
\[
(dL_{\beta}) = \left(c_q^{\beta} ~\prod_{i=1}^{m-1} q_i^{\beta-1} ~dq \right)~\left(m!~c_{L}^{\beta,a} e^{-\sum\limits_{i=1}^m \lambda_i/2} \Delta(\lambda)^{\beta} \prod_{i=1}^{m} \lambda_i^{a-p} ~d\lambda \right)~,
\]
where $c_q^{\beta}$ is the same as in (\ref{constq}).

{F}rom the above we see that $q$ and $\lambda$ are independent, and once
we drop the ordering the joint eigenvalue density is given by the
$\beta$-Laguerre ensemble of parameter $a$, while $q$ is distributed like a normalized vector of $\chi_{\beta}$'s. 

This concludes the proof of Theorem \ref{main2}.\\
\qed

\section{Applications and Open Problems} 

As we mentioned in Section \ref{intro}, we
believe that there should be many applications for the new tridiagonal
ensembles. We illustrate here some (in Section IV.A), in the hope that researchers will
find many more. Some of the applications we believe are new results (Applications 1, 3, 5, and 6), and some are simplifications of known results (Applications 2 and 4). 

We discuss the open problem of constructing a matrix model for the $\beta$-Jacobi ensembles in the beginning of Section IV.B. To facilitate the finding of new results, we conclude with a few open ``general $\beta$-ensemble'' problems.

\subsection{Applications}

\begin{enumerate}
\item \textit{Interpolating Laguerre exponents}. Our $\beta$-Laguerre ensembles have ``continuous'' Laguerre parameters $a$ which, even in the cases $\beta = 1,2,4$, interpolate the Wishart parameters. 

Though $\beta$-Laguerre ensembles with general (``continuous'') parameter $a$ have been studied by many researchers (\cite{Forrester_poly}, \cite{Johnstone}, \cite{muirhead82a}), no non-quantized matrix realizations (i.e. explicit random matrix models) of $\beta$-Laguerre ensembles are found in the literature. 

By ``quantized'' we mean that the exponent $a$ is either an even integer, an integer, or a half-integer (depending on the value of $\beta$). In particular, all models corresponding to a Laguerre (or Jacobi) weight found in \cite{Zirnbauer_10fold} and \cite{ivanov} are quantized.

Thus, our $\beta$-Laguerre random matrix constructions extend the pre-existing ones in two ways: through $\beta$ and through the Laguerre parameter $a$.

\item \textit{The expected characteristic polynomial}. The
result below might be seen as an extension of the classical Heine
theorem (see Szeg\"o \cite{szego} and Deift \cite{deift_book}) which
has $\beta = 2$. Note that for $\beta \neq 2$,
$\Delta(\lambda)^{\beta}$ can no longer be written as the determinant of a Vandermonde matrix times its transpose, and the proof cannot be duplicated. 

The same result is found in a slightly more general form in
\cite{edelman89a}, and its Jacobi case was first derived by Aomoto \cite{aomoto87a}.

\begin{theorem}
The expected characteristic polynomial $P_n(y)= $~{\rm det}$(yI_n-S)$ over $S$
in the $\beta$-Hermite, respectively $\beta$-Laguerre, ensembles 
are proportional to 
\[H_n \! \left(\frac{y}{\sqrt{2 \beta}}\right),~~~~~\mbox{respectively to}~~~~~ L_n^{ {\frac{2a}{\beta}-n}}\! \left(\frac{y}{2\beta} \right)~. 
\]
Here $H_n$ and $L_n^{{\frac{2a}{\beta}-n}}$ are the Hermite, respectively Laguerre,
polynomials, and the constant of proportionality accounts for the fact that $P_n(y)$ is
monic.
\end{theorem}

\begin{proof}
Both formulas follow immediately from the 3-term recurrence for the
characteristic polynomial of a tridiagonal matrix (see formula
(\ref{unu})) and from the independence of the variables involved in the 
recurrence. 
\end{proof}

\item \textit{Expected values of symmetric polynomials}. Using the three-term recurrence for the characteristic polynomial of a tridiagonal matrix, we obtain Theorem \ref{symmpol}.

\begin{theorem} \label{symmpol}
Let $p$ be any fixed (independent of $\beta$) multivariate symmetric polynomial on $n$ variables. Then the expected value of $p$ over the $\beta$-Hermite or $\beta$-Laguerre ensembles is a polynomial in $\beta$. 
\end{theorem}

We remark that it is difficult to see this from the eigenvalue density.

\begin{proof} The elementary symmetric functions 
\[
e_i(x_1, x_2, \ldots, x_n) = \sum_{1 \leq j_1<  \ldots <j_i \leq n } x_{j_1} x_{j_2} \ldots x_{j_i} ~~~~i = 0, 1, \ldots, n~,
\]
can be used to generate any symmetric polynomial of degree $n$ (in particular $p$). 

The $e_i$ evaluated at the eigenvalues of a matrix are the coefficients of its characteristic polynomial, and hence they can be written in terms of the matrix entries. Thus $p$ can be written as a polynomial of the $n \times n$ tridiagonal matrix entries (which corresponds to the Hermite, respectively Laguerre cases). 

To obtain the expected value of $p$ over the $\beta$-Hermite or $\beta$-Laguerre ensemble, one can write $p$ in terms of the corresponding matrix entries, use the symmetry to condense the expression, then replace the powers of the matrix entries by their expected values. 

The diagonal matrix entries are either normal random variables in the
Hermite case or sums of $\chi^2$ random variables in the Laguerre
case. The subdiagonal entries appear only raised at even powers in the $e_i$ and hence in $p$ (this is an immediate consequence of the three-term recurrence for the characteristic polynomial, (\ref{unu})). Since all even moments of the involved $\chi$ distributions are polynomials in $\beta/2$, it follows that the expectation of $p$ will be a polynomial in $\beta$.
\end{proof}

As an easy consequence we have the following corollary.

\begin{corollary}
All moments of the determinant of a $\beta$-Hermite matrix are \textbf{integer-coefficient} polynomials in $\beta/2$. 
\end{corollary}
\begin{proof}
Note that even moments of the $\chi_{\beta i}$ distribution are integer-coefficient polynomials in $\beta/2$, and that the determinant is $e_n$. 
\end{proof}

\item \textit{A new proof for Hermite and Laguerre forms of the
Selberg Integral}. Here is a quick proof for the Hermite and Laguerre
forms of the Selberg Integral (\cite{mehta_book}), using the $\beta$-Hermite,
respectively, $\beta$-Laguerre ensembles.

The Hermite Selberg integral is
\[
I_H(\beta, n) \equiv \int_{\mathbb{R}^n} |\Delta(\lambda)|^{\beta}
e^{-\sum_{i=1}^n \lambda_i^2/2} ~d\lambda
\]
We have that 
\[
 I_H(\beta, n)~ = ~ n! \left(\int_{0 \leq \lambda_1 \leq \ldots \leq \lambda_n< \infty} \Delta(\lambda)^{\beta}
e^{-\sum_{i=1}^n \lambda_i^2/2}~d\lambda \right)~~
\left(c_q^{\beta}\int_{S_{+}^{n-1}} \prod_{i=1}^n q_i^{\beta-1} ~dq \right)~,
\]
where $c_{q}^{\beta}$ is as in (\ref{constq}). We introduce the $n!$ because in the first integral we have ordered the eigenvalues; $S_{+}^{n-1}$ signifies that all $q_i$ are positive.

Note that $c_{q}^{\beta}$ can easily be computed independently of the $\beta$-Hermite ensembles. 

Using the formula for the Vandermonde given by Lemma \ref{one}, the
formula for the Jacobian $J$ given in Lemma \ref{two}, and
the fact that the Frobenius norm of a matrix in the tridiagonal $1$-Hermite
ensemble is the same as the Frobenius norm of its eigenvalue
matrix, one obtains
\begin{eqnarray*}
I_H(\beta, n) & = & n! ~c_{q}^{\beta} \int_{\mathbb{R}^{n} \times (0, \infty)^{n-1}} 
\frac{\prod_{i=1}^n q_i}{\prod_{i=1}^{n-1} b_i}
\frac{\prod_{i=1}^{n-1} b_i^{\beta i}}{\prod_{i=1}^{n}q_i^{\beta}} \prod_{i=1}^{n}q_i^{\beta-1}
e^{-\sum_{i=1}^n \beta_i^2 -\sum_{i=1}^n a_i^2/2}~da~db~\\
& = &  n! ~c_{q}^{\beta} (2 \pi)^{n/2} ~\prod_{i=1}^{n-1}\int_{(0, \infty)}b_i^{\beta i -1}
e^{-b_i^2} ~db_i~ \\
& = & n! ~\frac{2^{n-1}\Gamma (\frac{\beta}{2} n)}{\left(\Gamma (\frac{\beta}{2}) \right)^n} (2 \pi)^{n/2}~\prod_{i=1}^{n-1} \frac{\Gamma(\frac{\beta}{2}i)}{2}~~= ~~\frac{1}{c_{H}^{\beta}}~.
\end{eqnarray*}

The same reasoning yields the Laguerre Selberg Integral formula
\[
I_{L}^{\beta,a,n} = \frac{1}{c_{L}^{\beta,a}}~.
\]

\item \textit{Moments of the discriminant}. The discriminant of a polynomial equation of order $m$ is the square of the Vandermonde determinant of the $m$ zeroes of the equation. Thus, the discriminant of the characteristic polynomial of a $\beta$-Hermite of $\beta$-Laguerre ensemble matrix is simply $D(\lambda) = \Delta(\lambda)^2$.

A simple calculation shows that the $k$th moment of $D(\lambda)$ is 
\begin{eqnarray*}
\frac{c_{H}^{\beta}}{c_{H}^{\beta+2k}} & = & \prod_{j=1}^n 
\frac{(1+\frac{\beta}{2}j)_{kj}}{(1+\frac{\beta}{2})_{k}}~,  ~~~~~~\mbox{respectively,} \\
\frac{c_{L}^{\beta,a}}{c_{L}^{\beta+2k,~a+k(m-1)}}  & = & 2^{km(m-1)} \prod_{j=1}^m 
\frac{(1+\frac{\beta}{2}j)_{kj} ~(a-\frac{\beta}{2}(m-j))_{k(j-1)}}{(1+\frac{\beta}{2})_{k}}~.
\end{eqnarray*}

where $n$ and $m$ are the matrix sizes for the Hermite, respectively, Laguerre cases, and the rising factorial $(x)_{k} \equiv \Gamma(x+k)/\Gamma(x)$.

Using the Selberg integral, one obtains that the moments of the discriminant for the $\beta$-Jacobi case are
\begin{eqnarray*}
\frac{c_{J}^{\beta,~a_1,~a_2}}{c_{J}^{\beta +2k, ~a_1+k(m-1),~a2 +k(m-1)}} = \prod_{j=1}^m 
\frac{(1+\frac{\beta}{2}j)_{kj} ~(a_1-\frac{\beta}{2}(m-j))_{k(j-1)} ~(a_2-\frac{\beta}{2}(m-j))_{k(j-1)}}{(1+\frac{\beta}{2})_{k} ~(a_1+a_2 - \frac{\beta}{2}(m-j))_{k(m+j-2)}}~.
\end{eqnarray*}

\item \textit{Software for Application 2. Computing eigenvalue statistics for the $\beta$-ensembles.}
Application 2 suggests that integrals of the form 
\[
E_{\beta}[p] \equiv c_{H}^{\beta} \int_{\mathbb{R}^n} p(\lambda) ~|\Delta(\lambda)|^{\beta} ~e^{-\sum_{i=1}^n \lambda_i^2/2} ~d\lambda
\]
may be evaluated with software.

One example of this would be computing moments of the determinant over the $\beta$-Hermite ensemble. There are explicit formulas for the cases $\beta = 1, 2$ and
$4$, due to Mehta \cite{mehta_moments} and to Delannay and Le
Ca{\"e}r \cite{Del-Lec}, which can be used to evaluate these
moments.

In the absence of a closed-form, explicit formula, like the one for $\beta=1$
provided in \cite{Del-Lec}, the computation of these moments cannot be
made polynomial; thus it is inherently slow.

For the general $\beta$ case, one can compute the moments in terms of a multivariate Hermite polynomial evaluated at
$0$ (see \cite{LAA}, \cite{Forrester_poly}). Using this technique, the complexity of the computation exceeds by far that of symbolically taking the determinant of a tridiagonal matrix, expanding the power, and replacing all powers of the entries by their expected
values (which are all known). Writing a Mathematica code to implement this algorithm 
is an easy exercise, and such a code would allow the author to
compute these moments in a reasonable amount of time, provided that
the product between the power and the size of the matrix is not very
large.  A template for a special case when $\beta = 1$ can be found in \cite[Appendix A]{edelman97a}.

\end{enumerate}

\subsection{Open Problems}

\begin{enumerate}
\item \textit{$\beta$-Jacobi (MANOVA) Ensembles}.
Sections II and III of the paper provide tridiagonal matrix models for
the $\beta$-Hermite and $\beta$-Laguerre ensembles. The natural question is whether such models exist for the last member of the classical
triplet, Jacobi. The $\beta$-Jacobi
ensembles have been intensively studied as theoretical distributions,
especially in connection with Selberg-type integrals and Jack (or
Jack-Selberg) polynomials (see \cite{aomoto87a}, \cite{kadell_jacks}, 
\cite{kaneko}, \cite{barsky-carpentier}). Finding a random matrix
model that corresponds to them would be of much interest.

If the two matrix factorizations problems that are associated with the
Hermite and Laguerre ensembles are the EIG and the SVD, the one associated with the Jacobi should be the QZ
(the generalized symmetric eigenvalue problem). This idea is
supported by the fact that the MANOVA real and complex distributions,
which correspond to the Jacobi $\beta = 1,2$ ensembles, are indeed
connected to the QZ algorithm.
A good reference for QZ is \cite{golub-vanloan}. 

Though we have not studied this problem sufficiently, we believe that a concrete (perhaps sparse, perhaps tridiagonal) matrix model may be constructed for the $\beta$-Jacobi ensembles. 

\item \textit{Level densities.} 
The level density of an ensemble is the distribution of a random eigenvalue of that ensemble (and by the Wigner semicircular law we know that the limiting distribution as $n \rightarrow \infty$ of such an eigenvalue is semicircular). The three functions found to be the level densities of the Gaussian models depend on the univariate Hermite polynomials.

Recently, Forrester \cite{Forrester_book} has found a formula for the level densities of the $\beta$-Hermite ensembles which works for $\beta$ an even integer. This formula depends on a multivariate Hermite polynomial. 

Finding a unified formula for the general $\beta$ case would be of interest.

\item \textit{Level spacings.} The level spacings are the distances between the eigenvalues of an ensemble, usually normalized so that the average consecutive spacing is 1. These spacings have been well-studied in the case of the Gaussian ensembles ($\beta = 1,2,4$). The limiting probability density of a random spacing in these cases is known in terms of spheroidal functions (see \cite{mehta_book}).

A surprising connection exists between the limiting probability density of a GUE random spacing and the probability density of the zeroes of the Riemann zeta function. Inspired by the theoretical work of Montgomery (\cite{montgomery72}), Odlyzko (\cite{odlyzko01}) has shown experimentally that the two probability densities are very close; the subsequent conjecture that the two probability densities coincide has been named the Montgomery-Odlyzko law.

To the best of our knowledge, the level spacing of the general $\beta$-Hermite ensembles has not been investigated.

\item \textit{Bulk and edge scaling limits.} Finally, a very important application would be the generalization of the bulk and edge scaling limits for the GOE, GUE and GSE obtained by Tracy and Widom (the latter are known as the Tracy-Widom distributions $F_1$, $F_2$ and $F_4$). 

The edge scaling limit refers to the distribution of the largest eigenvalue of a matrix in the ensemble; the bulk scaling limit refers to the distribution
of an eigenvalue in the ``bulk'' of the spectrum. For a reference, see \cite{tracy_widom_universality}, \cite{tracy_widom_largest} or
\cite{tracy_widom_1_4}. The Tracy-Widom distributions are defined in terms of Painlev\'e functions, which are solutions to certain differential equations,
with asymptotics given by Airy functions. For a good treatment of Painlev\'e equations in relationship with Gaussian (Hermite), Laguerre, and Jacobi random
matrix models, see Pierre van Moerbeke's notes \cite[Section 4]{moerbeke}.  Recently, Johnstone \cite{Johnstone} has found that the limiting distributions $F_1$ and $F_2$ apply to real (respectively complex) Wishart matrices.

\end{enumerate}

\section{Acknowledgments}

The authors would like to thank Percy Deift, Peter Forrester, John Harnad, David Jackson, Eric Kostlan, Gene Shuman, Gil Strang, Harold Widom, and Martin Zirnbauer, for interesting and helpful conversations. We also thank the referee for challenging us to write a better paper, and for the many valuable writing suggestions.

Ioana Dumitriu's research was supported by an IBM PhD Fellowship and NSF grant DMS-9971591. Alan Edelman's research was supported by NSF grant DMS-9971591.

\pagebreak

\bibliography{bib_trid}
\bibliographystyle{plain}

\end{document}